\documentclass[twocolumn,english,showpacs,floatfix]{revtex4}
\usepackage{babel,amsmath,amssymb,dcolumn}
\usepackage{hyperref}
\usepackage{amsfonts}
\usepackage{graphicx}
\usepackage{latexsym}
\usepackage{dcolumn}

\newcommand{\be}{\begin{equation}}
\newcommand{\ee}{\end{equation}}

\newcommand{\bear}{\begin{eqnarray}}
\newcommand{\eear}{\end{eqnarray}}
\newcommand{\bearst}{\begin{eqnarray*}}
\newcommand{\eearst}{\end{eqnarray*}}
\begin{document}

\title{A Model for Dark Energy Decay}

\author{Elcio Abdalla}
\email{eabdalla@fma.if.usp.br}

\author{L. L. Graef}
\email{leilagraef@usp.br}

\affiliation{Instituto de F\'{\i}sica, Universidade de S\~ao Paulo, CP 66318, 05315-970, S\~ao Paulo, Brazil}

\author{Bin Wang}
\email{wang_b@sjtu.edu.cn}

\affiliation{INPAC and Department of Physics, Shanghai Jiao Tong University, 200240 Shanghai, China}

\begin{abstract}
We discuss a model of non perturbative decay of dark energy. We suggest the possibility that this model can provide a mechanism from the field theory to realize the energy transfer from dark energy into dark matter, which is the requirement to alleviate the coincidence problem. The advantage of the model is the fact that it accommodates a mean life compatible with the age of the universe. We also argue that supersymmetry is a natural set up, though not essential.
\end{abstract} 
\maketitle

\section{Introduction}

It is a rather accepted fact that our universe
contains about $70\%$ dark energy (DE), $25\%$
cold dark matter (DM) and a remaining fraction of
baryonic matter \cite{1}. In the concordance
model the cosmological constant is the easiest
explanation for the DE.  However, it is difficult
to understand the cosmological constant in terms
of fundamental physics. Its observed value is too
small, a fact referred to as the cosmological
constant problem. The fact that the amount of DE
and of DM are of the same order of magnitude
today is neither easy to comprehend, as we know
that in the past they differed by several orders
of magnitude. This is known as the coincidence
problem.

An interaction in the dark sector leads to a
mechanism to alleviate the coincidence problem
\cite{10}. Moreover, in the framework of field
theory, it is inevitable to consider an
interaction between DM and DE, given that they
are fundamental fields of the theory
\cite{secondref}. The dark sector interaction has
been widely discussed in the literature
\cite{xxx}-\cite{AbdallaPLB09}.  Extensive
analysis using the WMAP, SNIa, BAO and SDSS data
etc has been performed in refs. \cite{71}, as
well as the use of the late ISW effect to probe
the coupling between dark sectors \cite{hePRD09}.

A change in the growth index was found in refs. \cite{31,Caldera} as a consequence of the interaction, as well as further consequences concerning  the growth of cosmic structure \cite{31}-\cite{AbdallaPLB09}.  More recently, the interaction has been seen as an external potential leading to observable corrections to the Layser Irvine equation \cite{pt,AbdallaPLB09}. As a consequence, a small positive coupling has been tightly constrained \cite{AbdallaPLB09} in agreement with the results given in \cite{hePRD09} from CMB. The small positive coupling indicates that there is energy transfer from DE to DM, which helps alleviate the coincidence problem \cite{hePRD09,heJCAP08}.


Another possible explaination for the universe acceleration is  achieved in finding alternatives to the Einstein gravity. An example is the $f(R)$ gravity,  constructed based on a Lagrangian density given by an arbitrary function $f(R)$ depending on the curvature scalar \cite{fR}. $f(R)$ gravity is considered as the simplest modification to Einstein's general relativity.  The $f(R)$ gravity turns out to be conformally equivalent to an interaction model between DE and DM \cite{tsujikawa}. In the Einstein frame, the model does not possess a standard matter-dominated epoch as in the Jordan frame, but contains the coupling between the canonical scalar field to the non-relativistic matter. It was found that the condition that $f(R)$ gravity avoids a short-timescale instability and maintains the agreement with CMB is exactly equivalent to the requirement of an energy flow from DE to DM in the interaction model, which ensures the alleviation of the coincidence problem in the Einstein frame \cite{our2}.


\section{The Interacting Model}

When there is an energy exchange between dark
energy and dark matter, none of them is
separately conserved. In such a case the
conservation equations are written as \be
\label{conservacao1}
\rho'_{DE}+3H\rho_{DE}(1+w_{DE})=Q_{DE} , \ee \be
\rho'_{DM}+3H\rho_{DM}(1+w_{DM})=Q_{DM} , \ee
where prime denotes the derivative with respect
to conformal time, $w=p/\rho$ and $Q$ is the
interaction factor. We can see, through
eq.(\ref{conservacao1}), that we can define an
effective $w$ for the dark energy as
$w_{eff}=w-\frac{Q_{DE}}{3H\rho_{DE}}$, which
accounts for the interaction.

Supposing a decay from dark energy into dark
matter, it is natural to expect that
phenomenologically $Q$ is proportional to the
energy density of dark energy and to the decay
rate $\Gamma$, \be
Q_{DM}=-Q_{DE}=\Gamma\rho_{DE}, \ee where
$\Gamma>0$ indicates an energy flow from DE to
DM.  By defining $\Gamma$ we can find $Q_{DE}$ .
We can also write the coupling term in the form
of $Q_{DE}=-\gamma H \rho_{DE}$ , where $\gamma =
\Gamma /H$ . Integrating equation
(\ref{conservacao1}) we get the evolution for the
dark energy density \be \rho_{DE}=\rho_{DE0}
a^{-3(1+w_{eff})}, \ee where $w_{eff}=w +
\gamma/3$.

We know that the lifetime of the dark energy must
be of the order of the age of the universe. If it
was much more the effect of the interaction would
be negligible, and on the other hand, if it was
much less, the value of the dark energy density
should have been much higher in the past, and the
coupling term $Q_{DE}$, in this case, might not
have the small value predicted by the
observations. Moreover, as it's well known, the
standard  $\Lambda CDM$ model fits very well the
various observational results available, so it
would be nice to have a model whose dynamics
could approach the dynamics of the standard
model, in this case we expect a dark energy
density not much higher than $10^{-47}GeV^{4}$
even in the past.

We can see through eq.(\ref{conservacao1}) that,
with a coupling, each component does not conserve
separately anymore, they evolve correlated.
That's why it becomes possible to reproduce a
scaling solution of the kind \cite{coincid1} \be
\rho_{DE}=\rho_{DM} a^{\xi}, \ee where
$\xi=-3w_{eff}$.

For the cosmological constant case, being
$\rho_{DE}$ a constant, we have $\xi=3$, which
suffers the coincidence problem. When $\xi=0$,
the ratio $\rho_{DM}/\rho_{DE}=const$ and there
is no coincidence problem \cite{coincid2}. If
there is energy decay from dark energy to dark
matter, we can have  $\xi<3$, which can
accommodate longer period for the energy
densities of dark energy and dark matter to be
comparable so that to alleviate the coincidence
problem.

\section{A Model for the Decay}

Here we propose a further model for the
interaction. Suppose that a positive cosmological
constant (e.g. de Sitter like cosmology) is
modeled by a non zero scalar vacuum energy, and
that such a non zero vacuum energy density is
very small, $V_0\sim 10^{-47} $GeV$^4$. We
suppose a scalar with doubly degenerated energy
minima and a small breaking term to provide such
a small energy difference. This is indeed very
rare and generally unnatural except for a well
known case, that is if there is a symmetry
forcing the vacua to be equal and a
nonperturbative break of that symmetry. There
actually exists such a theoretical model. The
Wess Zumino model \cite{wesszumino} has a set of
degenerated bosonic vacua as a consequence of
supersymmetry which presumably is broken only non
perturbatively. We thus suppose this is the case
and consider a bosonic potential
\begin{equation}
V(\phi)=\vert 2m\phi -3\lambda \phi^2\vert^2 + Q(\phi) \equiv U(\phi) + Q(\phi),
\end{equation}
where $\phi =\varphi +iB$ is a complex scalar of
mass $m$ and coupling $\lambda$. The first term,
$U(\phi)$, corresponds to the bosonic sector of
the Wess-Zumino Lagrangian and $Q(\phi)$ is a
supersymmetry breaking term of power law type.
The term $Q(\phi)$ is adjusted so that we have
the cosmological constant value at the metastable
minimum. The exact form of the breaking term,
however, is not needed for the computations.

This potential has a set of zeros, in
$\varphi=0$, and at $\varphi=
\frac{2m}{3\lambda}$, if $B=0$. Let us suppose
that we have only one (uncharged) bosonic field,
so $B=0$. There is an interaction with a fermion
which for our calculations here is irrelevant. It
should become here clear that supersymmetry is
not a requirement for the present work, its just
a motivation for a potential with such a form.
The potential for this model is illustrated in
Fig.1.

\begin{figure}[h]
	\centering
		\includegraphics{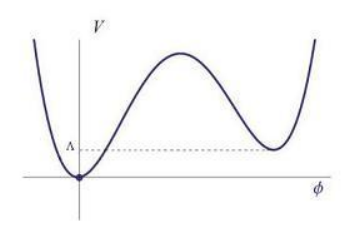}
	\caption{The Potential of the Field $\phi$.}
	\label{fig:figcap5}
\end{figure}

The physical mechanism is supposed to run as
follows. The field at the false vacuum represents
the dark energy. We know, however, that there can
be a decay from the false to the true vacuum.
After the field passes the potential barrier its
equation of state is no longer that of a dark
energy, and there is a non negligible kinetic energy. As happens in the old inflationary
scenario, the transition to the true vacuum
occurs through the formation of bubbles of new
vacuum. After these bubbles are nucleated, they
begin to expand very fast, as the energy of the
transition accelerates the bubble wall
\cite{coleman}. At some moment, however, the
walls of these bubbles, which carries the energy
of the transition, begin to collide. Through this
process, in the end, the energy released in the
conversion of the false vacuum into the true
vacuum can produce a kind of new component.

Since our field has negligible couplings to the
baryons (which is a reasonable supposition since
we have never detected such interaction), we
expect that the final product of the transition
must correspond to the dark sector. 
However, we are considering a supersymmetric potential, so we still have an interaction term  
$g\phi\bar{\psi}\psi$ (where $g$ is a coupling constant), which  
describes the talk between the scalar field and the fermionic field. So  
this new component produced can be pressureless fermionic cold dark  
matter. However since the decay time from the metastable  
vacuum to the stable vacuum of the scalar potential is of the order of the lifetime of the universe, the dark matter produced in $\phi$  
decay is not the dominant dark matter in the universe. In order for the standard  
cosmology to be recaptured, there has to be another dominant component  
of dark matter that was around high redshift, $z\sim 1000$.





In the following we will compute the decay from
the metastable vacuum to the stable vacuum of the scalar potential during
the lifetime of the universe. For certain values
of the parameters, the barrier has the exact
height needed for a decay occurring during this time.

\section{Computation of the Decay Rate}

The decay rate (per unit volume) from the metastable to the stable minimum of the potential $V(\varphi )$ is given, according to the semiclassical method, by \cite{coleman}
\be
\frac{\Gamma}{V}=\frac{S_{E}^{2}({\tilde{\varphi}}(\rho))}{(2\pi\hslash)^{2}}\times e^{-(\frac{S_{E}}{\hslash}-\frac{S_{\Lambda}}{\hslash})}\times
(\frac{det'(-\partial_{\mu}\partial_{\mu}+V''({\tilde{\varphi}}(\rho))}{det(-\partial_{\mu}\partial_{\mu}+V''(\varphi_{+}))})^{-\frac{1}{2}}
\ee
where $\varphi_{+}$ is the value of the field at
the false vaccum.  ${\tilde{\varphi}}(\rho)$ corresponds, in analogy with the case of particles, to the classical path in Euclidean space
crossing the potential  $-V(\varphi)$ with the boundary conditions
$\varphi_{initial}=\varphi_{final}=\varphi_{+}$. The
euclidean action $S_{E}$ in the above expression corresponds, in this
analogy, to the action of a particle in this oscilating trajectory,
and it is evaluated in relation to the action of a particle
at the false vacuum, $S_{\Lambda}$. The determinant is defined as
a ratio with respect to the determinant at the false vacuum, which
has the effect of a normalization.

The calculation of the decay rate is complicated but standard, a task
usually without any analytic solution. We use here the so
called thin wall approximation \cite{coleman}, in which the energy difference
between the two minima, given by the parameter $\epsilon$, is small,
and we can make the calculations perturbatively in $\epsilon$,
thus leading to an analytical solution.

The classical equation of motion, in the euclidean space, of the field $\varphi$ described
by the potential $V(\varphi )$, is obtained by minimizing the action
\bearst
\frac{\delta S_{E}(\varphi(x))} {\delta\varphi}=(-\partial_{\mu}\partial_{\mu}\varphi(x)+V'(\varphi))=0\quad .
\eearst

This is exactly the equation of motion of a scalar field in a potential
$-V(\varphi )$ in Minkowski space.

We suppose the boundary condition
\be
\underset{\tau\rightarrow{\pm}\infty}{lim}\varphi(\overset{\rightarrow}
{x},\tau)=\varphi_{+}.
\ee

Due to the symmetry of the problem we can assume the solution to be
euclidian invariant, thus, $\varphi(\overset{\rightarrow}{x},\tau)\rightarrow\varphi((\overset{\rightarrow}{|x|^{2}}+\tau^{2})^{\frac{1}{2}})$. For convenience we can define the variable $\rho=(|\overset{\rightarrow}{x}|^{2}+\tau^{2})^{\frac{1}{2}}$. In this case
the equation of motion becomes
\be \label{eqmotion}
\frac{\partial^{2}\varphi}{\partial\rho^{2}}+\frac{3}{\rho}\frac{\partial}{\partial\rho}\varphi -V'(\varphi)=0 \quad .
\ee
This equation of motion for the field
$\varphi$ is analog to the equation of motion of a particle at
position $\varphi$, moving in a time $\rho,$ subject to a potential
$-V(\varphi).$ The second term has a form analogous to a friction term. Observing the symmetry of the problem it is easy to see that the decay occurs by the formation of bubbles of true vacuum surrounded by the false vacuum outside. The term
$\frac{\partial\varphi}{\partial\rho}$ is  different from zero only
at the bubble wall, since the field is at rest inside and outside.
If this wall is thin we can consider $\rho = R $ in this region ($R$ denotes de radius of the bubble). When
$R$ is very large, as occurs when the energy
difference $\epsilon$ is small, we can neglect the friction term, as it is multiplied by $1/\rho$ that is equal to $1/R$ in the wall. So the equation
of motion becomes
\be
\frac{\partial^{2}\varphi}{\partial\rho^{2}}=V'(\varphi) \label{ddotphi}\quad .
\ee

The calculation of the action can be separated in three regions:
outside the bubble, at the thin wall and inside the bubble. In each of these regions the corresponding value of the field is
\begin{eqnarray*}
\varphi &=&2m/3\lambda\; , {\rm if } 0 <\rho\ll R \quad ,\\
\varphi &=&\tilde{\varphi}\; , \quad \mbox{   if}\quad  \rho\approx R\quad ,\\
\varphi &=&0\; , \quad  {\rm if} \quad \rho\gg R\quad .
\end{eqnarray*}
The exact form of  $\tilde\varphi$ is not needed here for our purposes. So, the expression for the action can be calculated as
\bear
&&S_{E}-S_{\Lambda}\approx  2\pi^{2}\intop_{0}^{R-\Delta}d\rho\rho^{3}(-\epsilon)+\nonumber\\
&&2\pi^{2}\intop_{R-\Delta}^{R+\Delta}d\rho\rho^{3}(\frac{1}{2}(\frac{d\tilde{\varphi}}{d\rho})^{2}+U)+2\pi^{2}\intop_{R+\Delta}^{\infty}d\rho\rho^{3}(0)\;,
\eear
where $\Delta$ represents the width of the wall.
After integrating we obtain the action
\bear
S_{E}-S_{\Lambda}&=&-2\pi^{2}\epsilon\frac{R^{4}}{4}+2\pi^{2}R^{3}\intop_{R-\Delta}^{R+\Delta}d\rho(\frac{1}{2}(\frac{d\tilde{\varphi}}{d\rho})^{2}+U)+0\nonumber\\
&=&-\frac{1}{2}\pi^{2}R^{4}\epsilon+2\pi^{2}R^{3}S_{1}\label{sdeR}\quad ,
\eear
where we defined $(S_{1}=\intop_{R-\Delta}^{R+\Delta}d\rho(\frac{1}{2}(\frac{d\tilde{\varphi}}{d\rho})^{2}+U))$.
From now on we call $S_{E}-S_{\Lambda}$ simply
as $S$.  We can find $R$ by minimizing the action
\be
\frac{dS}{dR}=-2\pi^{2}R^{3}\epsilon+6\pi^{2}R^{2}S_{1}=0\quad ,
\ee
and so we obtain $R=3S_{1}/\epsilon$.
We can see that $R\rightarrow\infty$ if $\epsilon\rightarrow0$.
This is the reason why in our approximation we neglected the friction term when
$\epsilon$ is very small.

Integrating equation (\ref{ddotphi}), considering
$\epsilon$ small, we obtain the relation $\frac{\partial}{\partial\rho}\varphi=\sqrt{2U}$, allowing
to get for $S_{1}$ the expression
\be
S_{1}=\sqrt{2}\int_{\varphi_{-}}^{\varphi_{+}}d\varphi\sqrt{U}\quad .
\ee





Substituting the expression of the Wess-Zumino supersymmetric potential
into the expression of $S_{1}$ we get
\be
S_{1}=\sqrt{2}\{\frac{4m^{3}}{27\lambda^{2}}\}\quad .
\ee

Using above expression in eq.(\ref{sdeR}) we can see that the action will have the form
\be
S\approx\frac{m^{12}}{\lambda^{8}\epsilon^{3}}.
\ee

Since the exponential term in the decay rate dominates whenever we
are within the validity of the semiclassical limit, the pre-exponential
term will change the result so insignificantly on the scale we are
working, that a simple estimate of it's order of magnitude is enough.
We know that the dimension of the pre-exponential term is of $m^{4}$
and it's value is determined by the parameters of the theory with dimension of
mass. Therefore we can estimate that for the energy scale we are dealing with, the decay rate per unit volume can be written as  \cite{19}
\be
\frac{\Gamma}{V}= m^{4} e^{-m^{12}/(\lambda^{8}\epsilon^{3})}.
\ee

We will simply estimate the pre-exponential terms
as 1 GeV$^{4}$ in order to facilitate the calculations and give the
correct units we are dealing with. Its easy to show that it will not affect our results.

So, by substituting the value of $\epsilon$ and $\lambda$, we obtain for the decay rate (per volume)
\be
\frac{\Gamma}{V}=e^{-10^{156} (\frac{m}{GeV})^{12}}GeV^{4}\quad .
\ee
Inverting the expression of the decay rate we then obtain the decay
time (times the volume) of a particle. Due to the symmetry of our problem we can simply take the fourth root of the result in order to obtain the following decay time
\be
t_{decay}=10^{-25}\{exp(10^{156}(\frac{m}{GeV})^{12})\}^{\frac{1}{4}}s.
\ee

If we equate this decay time to the age of the universe, ($10^{17}s$) we obtain for the mass the value
\be
m\sim10^{-13}GeV.
\ee

Having this result its easy to calculate that the value of R (the radius of the bubbles in the moment they are formed) is about $10^{-3}cm$. After the tunneling the field evolves according to the classical field equation, which is simply the analitic continuation of the euclidean field equation (\ref{eqmotion}) back to real time.

Similar results are obtained when using other potential
with similar characteristics, as for example $U=\frac{\lambda}{8}(\phi^{2}-\frac{m^{2}}{\lambda})^{2}$
plus a term that generates a metastable minimum with the energy density
of the cosmological constant. Using other potential of this kind we
obtain as result approximately the same order of magnitude for the mass.

If we consider in our calculations the gravitational effect we must consider the following action
\be
\mbox{\ensuremath{\bar{S}=\int d^{4}x$$\sqrt{-g}(\frac{1}{2}g^{\mu\nu}\partial_{\mu}\varphi\partial_{\nu}\varphi-V(\varphi)-\frac{R}{16\pi G})}}\quad .
\ee

It is possible to show that using the thin wall approximation we get the following relation between the action including gravity and the one we have calculated before \cite{18} ,
\be
\mbox{\ensuremath{\bar{S}=\frac{S_{0}}{(1+(\frac{R_{0}}{2\Delta})^{2})^{2}}}}\quad ,\label{17}
\ee
where ${S_{0}}$ is the expression for the action obtained in the previous case, ${R_{0}}$ is the radius of the bubble formed in that case, $\bar{S}$ the new action considering gravity and $\Delta$ is the Schwarzschild radius associated to the bubble of new vacuum.

In the case that gravity is included the thin wall approximation is a good approximation for all cases of our interest. It's possible to show that in this case we can neglect the friction term and also the expansion of the universe.

The energy released in the conversion of false into true vacuum is
proportional to the volume of the bubble of new vacuum formed. There
is therefore a Schwarzschild radius associated to this energy. We can see by the equation above that if the radius of the bubble of true vacuum is comparable to its Schwarzschild
radius then it is important to include gravity in our calculation
\cite{18}, otherwise we recover our previous result.
Equating the radius $R$ of our bubble to the expression of the
Schwarzschild radius it's easy to show that, for the scale of the energies we are working with, we really could have simply neglected the gravitational effects.

Let us briefly discuss the symmetry breaking that
generates the false vacuum energy density. As $\varphi\sim\frac{m}{\lambda}$ at this
point, we can see that terms such as
\be
Q(\varphi)=m^{2}\varphi^{2},\lambda m\varphi^{3},\lambda^{2}\varphi^{4},
\ee
breaks the symmetry and causes this vacuum to
have the energy density corresponding to the cosmological constant.

\section{Conclusion}

We calculated the decay of a particle of dark energy, with mass of the order
 $m\sim10^{-13}$GeV, described by the potential
$V(\varphi)=|2m\varphi-3\lambda\varphi^{2}|^{2}+Q(\varphi)$ , from the metastable to the stable minimum of the potential. The timescale of such a process is compatible with the order of magnitude of the age of the universe. We suggest that such a kind of quintessence model can provide a mechanism, from field theory, to explain the decay of dark energy into dark matter, alleviating the coincidence problem of the concordance model of cosmology.

We think that, in a future work, a further analysis of the evolution of the bubbles of new vacuum after its formation, could give us previsions for the size and configuration of these bubbles, which could be a potentially testable signal of this model.

In view of the arguments in \cite{our2}, our model can be equivalent to an $f(R)$ gravity, with a calculable function $f(R)$. We also think it is interesting to investigate the $f(R)$ gravity based on the proposed field theory model.

\acknowledgements{The authors wish to thank CNPq and FAPESP, Brazil and NNSF, China, for support.}

\end{document}